\documentstyle{mn}

\newif\ifAMStwofonts
\AMStwofontstrue

\input{epsf}

\def\ginga{{\it Ginga}}
\def\asca{{\it ASCA}}
\def\wfe{W_{\rm K\alpha}}
\def\efe{E_{\rm K\alpha}}
\def\sfe{\sigma_{\rm K\alpha}}
\newcommand{\sax}{{\it Beppo\-SAX}}
\newcommand{\xte}{{\it RXTE}}

\ifoldfss
  \ifCUPmtlplainloaded \else
    \NewTextAlphabet{textbfit} {cmbxti10} {}
    \NewTextAlphabet{textbfss} {cmssbx10} {}
    \NewMathAlphabet{mathbfit} {cmbxti10} {} 
    \NewMathAlphabet{mathbfss} {cmssbx10} {} 
  \fi
  \ifAMStwofonts
    \ifCUPmtlplainloaded \else
      \NewSymbolFont{upmath} {eurm10}
      \NewSymbolFont{AMSa} {msam10}
      \NewMathSymbol{\upi}     {0}{upmath}{19}
      \NewMathSymbol{\umu}     {0}{upmath}{16}
      \NewMathSymbol{\upartial}{0}{upmath}{40}
      \NewMathSymbol{\leqslant}{3}{AMSa}{36}
      \NewMathSymbol{\geqslant}{3}{AMSa}{3E}

      \let\geq=\geqslant 
    \fi
  \fi
\fi 

\ifnfssone
  \newmathalphabet{\mathit}
  \addtoversion{normal}{\mathit}{cmr}{m}{it}
  \addtoversion{bold}{\mathit}{cmr}{bx}{it}
  \newmathalphabet{\mathbfit} 
  \addtoversion{normal}{\mathbfit}{cmr}{bx}{it}
  \addtoversion{bold}{\mathbfit}{cmr}{bx}{it}
  \newmathalphabet{\mathbfss} 
  \addtoversion{normal}{\mathbfss}{cmss}{bx}{n}
  \addtoversion{bold}{\mathbfss}{cmss}{bx}{n}
  \ifAMStwofonts
    \ifCUPmtlplainloaded \else
      \UseAMStwoboldmath
      \makeatletter
      \new@mathgroup\upmath@group
      \define@mathgroup\mv@normal\upmath@group{eur}{m}{n}
      \define@mathgroup\mv@bold\upmath@group{eur}{b}{n}
      \edef\UPM{\hexnumber\upmath@group}
      \new@mathgroup\amsa@group
      \define@mathgroup\mv@normal\amsa@group{msa}{m}{n}
      \define@mathgroup\mv@bold\amsa@group{msa}{m}{n}
      \edef\AMSa{\hexnumber\amsa@group}
      \makeatother
      \mathchardef\upi="0\UPM19
      \mathchardef\umu="0\UPM16
      \mathchardef\upartial="0\UPM40
      \mathchardef\leqslant="3\AMSa36
      \mathchardef\geqslant="3\AMSa3E

      \let\geq=\geqslant 
    \fi
  \fi
\fi 

\ifnfsstwo
  \DeclareMathAlphabet{\mathbfit}{OT1}{cmr}{bx}{it}
  \SetMathAlphabet\mathbfit{bold}{OT1}{cmr}{bx}{it}
  \DeclareMathAlphabet{\mathbfss}{OT1}{cmss}{bx}{n}
  \SetMathAlphabet\mathbfss{bold}{OT1}{cmss}{bx}{n}
  \ifAMStwofonts
    \ifCUPmtlplainloaded \else
      \DeclareSymbolFont{UPM}{U}{eur}{m}{n}
      \SetSymbolFont{UPM}{bold}{U}{eur}{b}{n}
      \DeclareSymbolFont{AMSa}{U}{msa}{m}{n}
      \DeclareMathSymbol{\upi}{0}{UPM}{"19}
      \DeclareMathSymbol{\umu}{0}{UPM}{"16}
      \DeclareMathSymbol{\upartial}{0}{UPM}{"40}
      \DeclareMathSymbol{\leqslant}{3}{AMSa}{"36}
      \DeclareMathSymbol{\geqslant}{3}{AMSa}{"3E}

      \let\geq=\geqslant 
    \fi
  \fi
\fi 

\ifCUPmtlplainloaded \else
  \ifAMStwofonts \else 
    \def\upi{\upi}
    \def\umu{\mu}
    \def\upartial{\partial}
  \fi
\fi

\title[The strength and width of Fe K$\alpha$ lines in Seyferts]
{The strength and width of Fe K$\bmath{\alpha}$ lines in Seyferts and their
correlations with the X-ray slope}

\author[P. Lubi\'nski and A. A. Zdziarski]
{\parbox[]{7in} {Piotr Lubi\'nski$^{1,2}$\thanks{E-mail: piotr@camk.edu.pl} 
and Andrzej A. Zdziarski$^1$}\\
$^1$N. Copernicus Astronomical Center, Bartycka 18, 00-716 Warsaw, Poland \\
$^2$Heavy Ion Laboratory, Warsaw University, Pasteura 5a, 02-093 Warsaw, Poland
\\
}

\date{Accepted 2001 March 8. Received 2001 February 28; in original form
2000 September 6}

\pagerange{L37--L42}
\volume{323}
\pubyear{2001}

\begin{document}

\maketitle

\label{firstpage}

\setcounter{page}{37}

\begin{abstract} We study properties of Fe K lines of a large sample of Seyfert 
1s observed by \asca. Fits with power laws and Gaussian lines yield the average 
linewidth and equivalent width of $ 0.22\pm 0.03$ keV and $0.13\pm 0.01$ keV, 
respectively. Thus, the typical lines are weak and narrow. We then 
obtain the average line profile of all our spectra, and find it to consist of a 
narrow core and blue and red wings, with the red wing being much weaker than 
that of, e.g., MCG --6-30-15. We obtain three average spectra of Seyferts grouped 
according to the hardness, and find the equivalent width of the core 
(originating in a remote medium) to be $\simeq 50$ eV in all three cases. The wings 
are well fitted by a broad line from a disc with strong relativistic effects. 
Its equivalent width correlates with the slope, increasing from $\sim 70$ eV 
for the hardest spectrum to $\sim 120$ eV for the softest one. The inner disc 
radius decreases correspondingly from $\sim 40$ to $\sim 10$ gravitational 
radii, and the fitted disc inclination is $\sim 45\degr$. The obtained 
correlation between the slope and the strength of the broad Fe K line is found 
to be consistent with the previously found correlation of the slope and Compton 
reflection.  \end{abstract}


\section{INTRODUCTION}
\label{s:intro}

There are two, related, main X-ray signatures of the presence of cold matter in
accreting compact sources: Fe K$\alpha$ flouorescence line and Compton
reflection. The latter was discovered in Seyfert 1s by \ginga\/ (Pounds et al.\
1990), and the average solid angle, $\Omega$, subtended by the reflecting
medium was found to be $\Omega/2\upi\simeq 0.5$ (Nandra \& Pounds 1994). The
former has been observed in most detail so far by \asca, for which observations of
Seyfert 1s yielded typical Fe K$\alpha$ equivalent width, $\wfe$, of $\sim
200$--300 eV (Nandra et al.\ 1997, hereafter N97).

There is a clear discrepancy between these two results if the line is supposed to 
come from the same medium as the Compton-reflecting continuum. Namely, the line 
is formed following about a half of ionizations of the innermost electron shell 
of an Fe atom by photons above the K edge (at 7.1 keV for neutral iron) 
incident on the reflecting medium. This gives a relationship between $\Omega$ 
and $\wfe$, which, for the average photon index of Seyfert 1s of $\Gamma \simeq 
1.95$ (Nandra \& Pounds 1994) and the viewing angle of $i=30\degr$ (N97) gives 
$\wfe/(\Omega/2\upi) \simeq (100$--130) eV, where the uncertainty of the 
coefficient is the result of the uncertain Fe K cross section (\.Zycki \& Czerny 1994; 
George \& Fabian 1991). Thus, the measured reflection implies the $\wfe$ of 
only 50--70 eV on average, much lower than that measured by N97.

A possible resolution of this dilemma is to postulate that the bulk of the line 
comes from a Thomson-thin molecular torus surrounding the 
X-ray source, which, for some geometries, can yield the required $\wfe$ 
without the reflection signatures (e.g.\ Wo\'zniak et al.\ 1998). However, 
the lines observed by \asca\/ are clearly broad (e.g., N97), which rules out 
their origin mostly from the torus. The most straightforward explanation of 
their width is a relativistic motion, presumably of an optically-thick 
accretion disc in the gravitational potential of the nucleus (Fabian et al.\ 
1989; Tanaka et al.\ 1995), in which case the presence of reflection is 
unavoidable. Then, models in which the line strength is increased due to 
relativistic effects (Martocchia \& Matt 1996; Reynolds \& Fabian 1997) would 
imply a corresponding enhancement of the reflection, which is not seen. This 
problem has apparently motivated some authors to distinguish ``reflection 
models" from disc models without reflection. This is clearly unphysical, since, 
as noted by N97, ``simple atomic physics dictates that the fluorescence must be 
accompanied by a Compton-scattered continuum".

Moreover, the strength of Compton reflection in Seyfert 1s has been found to 
correlate strongly with $\Gamma$ (Zdziarski, Lubi\'nski \& Smith 1999, 
hereafter ZLS99). This correlation has also been found in black-hole binaries, 
in which case $\wfe\propto \Omega$, as expected (Gilfanov, Churazov \& 
Revnivtsev 1999, 2001). On the other hand, no such proportionality has been 
claimed yet for Seyfert 1s, and some examples to the contrary have been seen 
(e.g.\ Chiang et al.\ 2000).

Here we re-examine the issue of the strength and width of the Fe K$\alpha$ line
in Seyferts and their correlation with the spectral index. We use the large
available database of \asca, which, at this time, contains about 3 times as
many observations as those available at the time of the study of N97. Moreover,
we make use of the \asca\/ calibration improved significantly over the
years of the mission.

\section{SIMPLE FITS TO INDIVIDUAL SPECTRA}
\label{s:individual}

Our Seyfert sample consists of the objects studied by N97 and by ZLS99 (except 
for NGC 4151 and NGC 6814, see below). These objects are mostly classified as 
radio-quiet Seyfert 1s or narrow emission-line galaxies. The latter are 
Seyferts intermediate between type 1 and 2 showing moderate X-ray absorption 
(e.g.\ Smith \& Done 1996). We include 3C 120, as it is present in the sample 
of N97. Although it is radio-loud, its Fe K$\alpha$ and reflection spectral 
features are similar to those of radio-quiet Seyferts (Zdziarski \& Grandi 
2001). We exclude NGC 4151, the X-ray brightest radio-quiet Seyfert, since it 
is characterized by an absorbing column, $N_{\rm H}\ga 10^{23}$ cm$^{-2}$, much 
larger than those of other objects in the sample. We exclude NGC 6814 due to 
the poor statistics of its spectrum.

We use \asca\/ spectra extracted from the public Tartarus 1 database. After
excluding 3 short observations with poor statistics, we have 71 spectra of 25
AGNs. We distinguish between 16 AGNs in the sample of N97: 3C 120,
Fairall 9, IC 4329A, MCG --2-58-22, MCG --6-30-15, Mrk 335, Mrk 509, Mrk 766,
Mrk 841, NGC 3227, NGC 3516, NGC 3783, NGC 4051, NGC 5548, NGC 7213, and NGC
7469 (58 observations); and 9 AGNs present only in the sample of ZLS99: AKN
120, MCG --5-23-16, NGC 2110, NGC 2992, NGC 4593, NGC 526A, NGC 5506, NGC 7172,
and NGC 7314 (13 observations). The log of observations and results of
individual fits are given in Lubi\'nski, Zdziarski \& Madejski (2001, hereafter
LZM01).

For fitting the Fe K lines  (using {\sc xspec}, Arnaud 1996), we use only the
3--10 keV continuum in order to avoid the complexities of possible soft X-ray
excesses and warm absorbers (as in N97). In those fits, we use the Galactic
$N_{\rm H}$ as well as $N_{\rm H}$ at the redshift of the source (assuming a
neutral medium with the abundances of Anders \& Ebihara 1982) determined by
modeling the entire \asca\/ spectrum by either a power law or a broken power
law. However, the effect of $N_{\rm H}$ on the 3--10 keV spectrum is minor. We
then model the spectrum as a power law and a sum of 3 Gaussian  lines. The
first component is the main Fe K$\alpha$ line, and the second is due to its
reflection from the cold  medium (George \& Fabian 1991), with the energy and
the width of 0.975 and the intensity of 0.12 of the main line. The third
component represents combined Fe K$\beta$ and Ni K$\alpha$ emission, with the
energy and the width of 1.125 and the intensity of 0.18 of the main line. Due
to limited statistics at $\ga 7$ keV, individual \asca\/ spectra do not allow
us to fit the Compton reflection continuum, which we thus neglect at this
stage. We fit the same model to the SIS0, SIS1, GIS2 and GIS3 data
allowing for free relative normalization with respect to SIS0.

\begin{figure}
\begin{center} \leavevmode \epsfxsize=8.4cm
\epsfbox{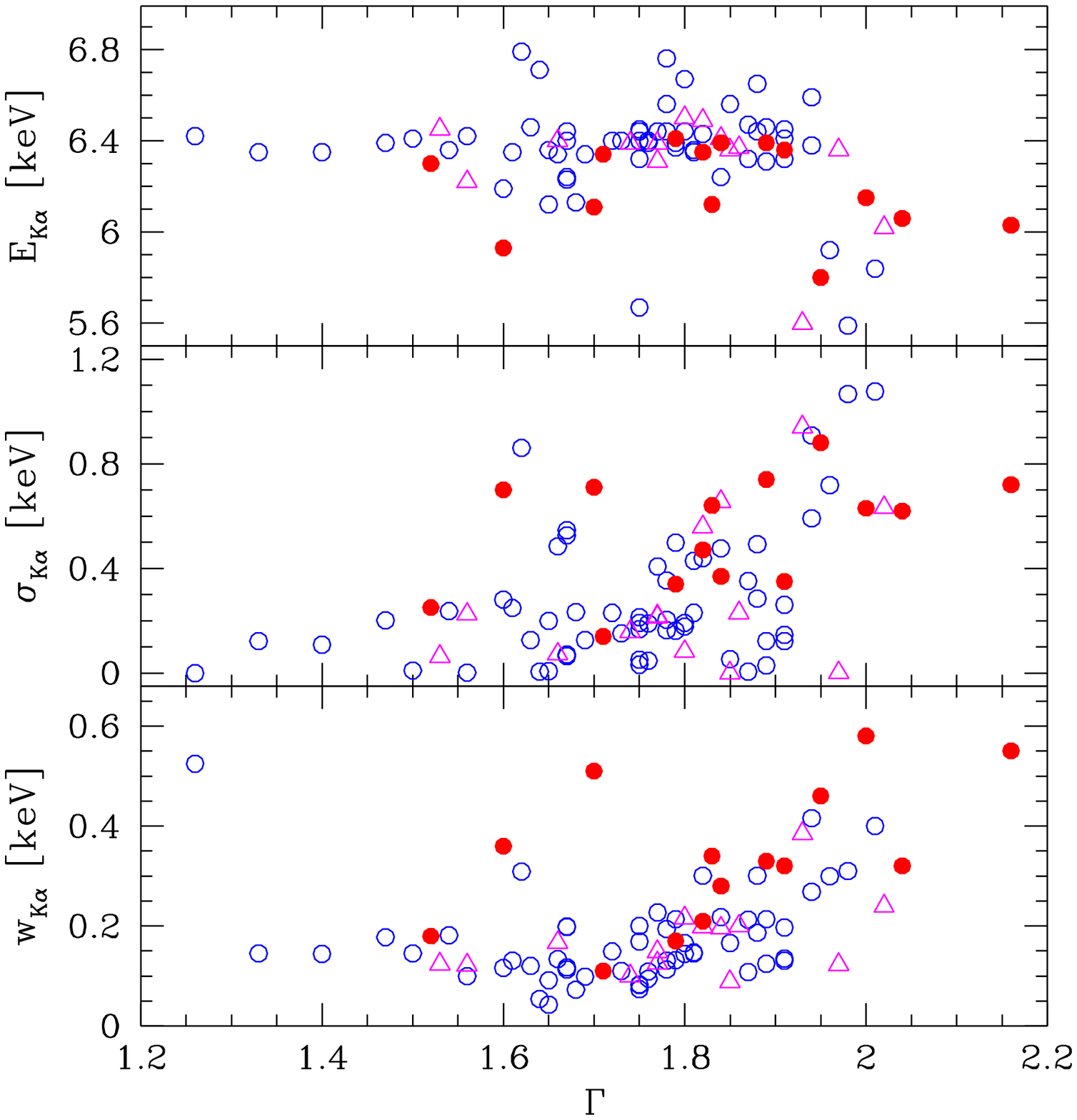}\end{center} \caption{The peak energy (top), the width
(middle) and the equivalent width (bottom) of Fe K$\alpha$ lines in Seyferts
observed by \asca\/ as functions of their 3--10 keV power-law index. The open
circles and filled circles represent our results and those of N97,
respectively,  for the same AGN sample. The open triangles show our results for
AGNs included in our sample but not in N97.
}
\label{fig:individual}
\end{figure}

Fig.\ \ref{fig:individual} shows the resulting parameters of the main Fe
K$\alpha$ line. We see that the lines are relatively weak and narrow, and 
peak close to 6.4 keV. The values of the weighted mean and its
$1\sigma$ error are $\langle \efe\rangle =6.39\pm 0.01$ keV, $\langle
\sfe\rangle = 0.22\pm 0.03$ keV, $\langle \wfe\rangle= 0.13\pm 0.01$ keV, and
$\langle \Gamma\rangle=1.75\pm 0.02$. We see no statistically significant
differences between the AGN sample of N97 and the objects outside it.

For comparison, Fig.\ \ref{fig:individual} also shows results of analogous fits 
of N97, where the weighted means are $\langle \efe\rangle =6.28\pm 0.04$ 
keV, $\langle \sfe\rangle = 0.37\pm 0.06$ keV, and $\langle \wfe\rangle= 
0.19\pm 0.03$ keV for the objects included here. 
We note that N97 used a single Gaussian in 
their fits, which, as we have checked, would have increased our values of 
$\sfe$ and $\wfe$ by about 20 per cent. We note that N97 obtained 
$\langle \sfe\rangle$ about twice that found here as well as their measurements 
show a scatter significantly larger than in our case, with a relatively large 
number of objects with $\sfe$ and $\wfe$ much above their respective mean 
values. The fact that the extreme data points have a little effect on the 
weighted means is due to the large errors for most objects with very broad 
lines in their sample.

The above differences are fully explained by improvements in the \asca\/ 
calibration since the analysis of N97, who used the calibration files of 1995. 
An illustrative example is that of 3C 120. N97 obtained $\sfe= 
0.74_{-0.27}^{+0.34}$ keV and $\wfe= 0.33_{-0.13}^{+0.20}$ keV (their error 
ranges are for $\Delta\chi^2=4.7$). Then, Wo\'zniak et al.\ (1998) obtained 
$\sfe=0.28^{+0.96}_{-0.16}$ keV (hereafter the errors 
are for 90 per cent confidence, i.e., $\Delta\chi^2=2.7$) and 
$\wfe\simeq 0.1$ keV, i.e., a line $\sim 3$ times weaker and narrower, using 
{\it the same data and model\/} as N97, but with the calibration of 1997. The 
weakness of the line has then been confirmed by broad-band observations by 
\sax\/ and  {\it RXTE}, which yielded $\sfe= 0.40_{-0.18}^{+0.87}$ keV, 
$0.09_{-0.09}^{+0.32}$ keV, and $\wfe= 0.08_{- 0.03}^{+0.07}$ keV, $0.09\pm 
0.02$ keV, respectively (Zdziarski \& Grandi 2001). 

\section{THE AVERAGE SPECTRUM}
\label{s:average}

Due to the limited effective area of \asca, the detailed shape of individual 
spectra of Seyferts is not well constrained, but the average form can be 
measured in much more detail. For that, we use only data from the SIS 
detectors, as they have a much better energy resolution than the GIS ones. We 
follow here the method of N97. We fit the SIS data of a single observation by a 
power law in the ranges of 3--4.5 keV and 7.5--10 keV, i.e., excluding the Fe K 
line region, obtain the ratio of the observed counts to those predicted by the 
power-law model, and then transform it to the rest frame.

Then, we obtain a weighted average of the ratios. A contribution to a bin in
the average ratio is weighted by both its inverse-square error and the ratio of
the length of the part of the input bin contributing to an output bin with
respect to the length of the output bin. The latter expresses conservation of
photons between the originally-binned spectrum and that with arbitrarily chosen
final bins.

\begin{figure}
\begin{center} \leavevmode \epsfxsize=8.4cm \epsfbox{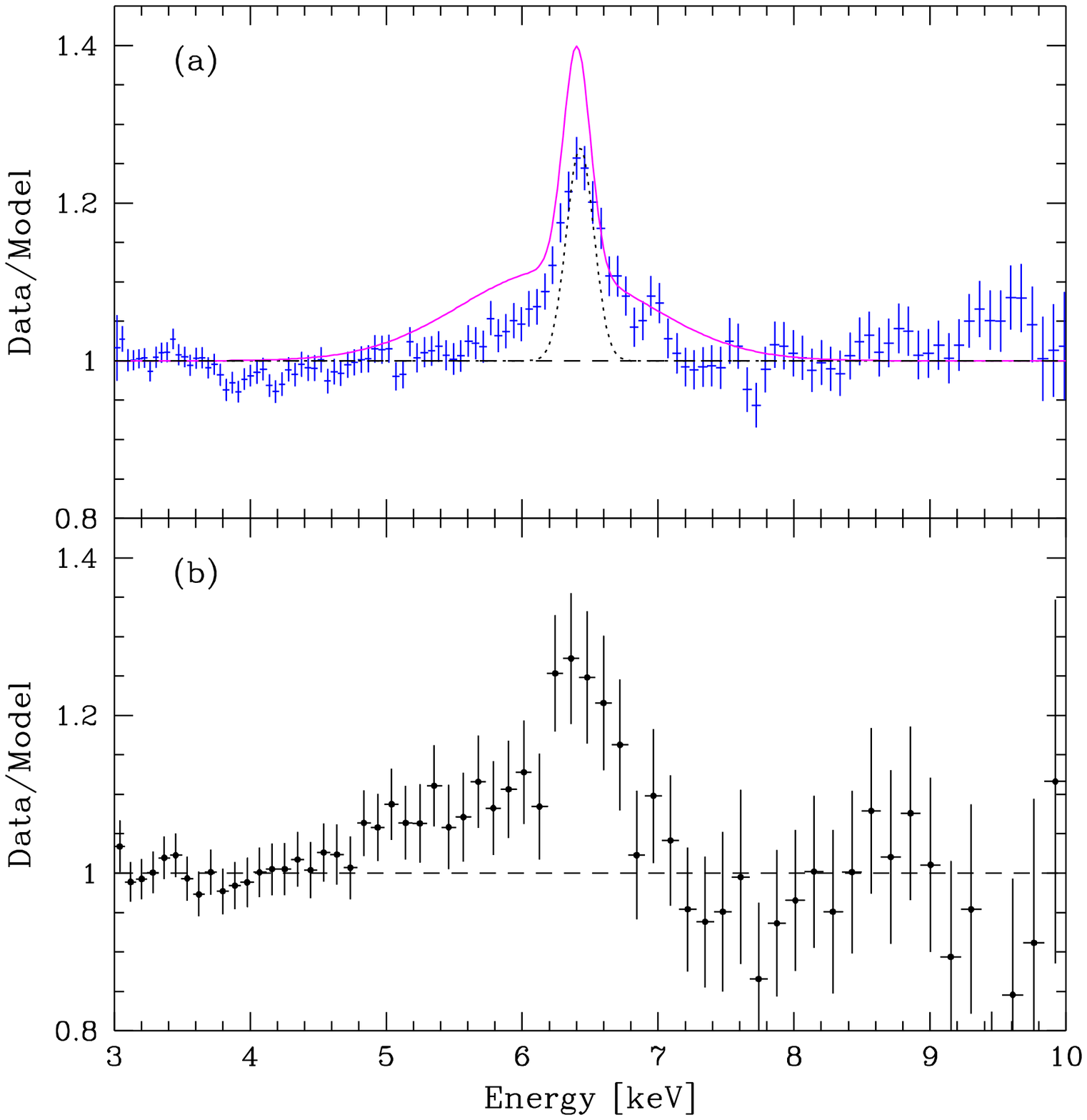} \end{center}
\caption{(a) The average line profile of the sample of Seyferts of N97 (except
NGC 4151 and NGC 6814, see Section 2). For comparison, the solid curve shows
the results of a double-Gaussian fit to the average profile of N97. The dotted
curve shows the response of \asca\/ to an intrinsically narrow line. (b) The
line profile averaged over \asca\/ observations of MCG --6-30-15. }
\label{fig:average}
\end{figure}

Fig.\ \ref{fig:average}(a) shows the obtained average for all observations of
the 16 AGNs of N97 (to improve statistics, we use all available observations
of those AGNs). Note that our method yields the average line profile in the SIS
count space (as in N97), and thus an intrinsically narrow line would have a
$\sfe^{\rm n} \simeq 0.1$ keV (corresponding to the SIS resolution averaged
over time), as shown by the dotted curve on Fig.\ \ref{fig:average}(a). The
observed profile clearly contains such a narrow core and relatively weak red
and blue wings. Both the red wing and the core are weaker than those obtained
by N97 (solid curve, obtained by scanning fig.\ 4a and normalizing using
fig.\ 5a in N97). We find that adding the 13 observations of the 9 objects not
present in the sample of N97 has virtually no effect on the obtained profile.

We have found that the main factor explaining the above discrepancy 
appears to be 
the change of the \asca\/ calibration in 1996 (see LZM01 for details). We have 
obtained this result by dividing the line profile of N97 by the 
curve\footnote{from http://heasarc.gsfc.nasa.gov/docs/asca/\hfill\\ 
xrt$\_$new$\_$response$\_$announce/announce.html} by which the effective area 
of \asca\/ was corrected in 1996.  After a renormalization, the obtained shape 
is close to our average profile. A further factor is the inclusion of NGC 4151 
in the profile of N97. The brightness of this object causes it to weigh heavily 
in the average, and to increase noticeably the red wing of the line. However, 
the very strong red wing of the line in this object obtained with fits with a 
single absorber may be spurious (Zdziarski, Johnson \& Magdziarz 1996). 
We also find that limiting the sample to the 19 early 
observations of the 16 AGNs included in N97 has virtually no effect on the 
strength of the red wing. On the other hand, the core of that spectrum is 
narrower, which is due to the time-dependence of the \asca\/ resolution. 

We have extensively tested our results. First, we obtain a good agreement with 
results for the objects with confirmed broad lines with red wings. Fig.\ 
\ref{fig:average}(b) shows our obtained average profile of MCG --6-30-15, which 
does have a very pronounced red wing, in agreement with previous results 
(Tanaka et al.\ 1995; Iwasawa et al.\ 1996). Then, we have have reproduced very 
closely the results of Nandra et al.\ (1999) for another object with a very 
strong and broad line, NGC 3516. Also, we have studied the subsamples of SIS0 
and SIS1 separately, and also the subsamples of the observations made before and after
the end of 1995. We have found no significant differences between various subsamples, 
except between early and late \asca\/ spectra, with the latter spectrum being 
broader due to the well-known effect of degradation of the SIS resolution with 
time. Finally, we have studied 10 observations for which we also used the 
spectra from the Tartarus 2 database. We have found only very tiny differences 
in individual fits. Their average profiles obtained the 2 databases show very 
good agreement as well as neither shows a strong red wing, in agreement with 
Fig.\ \ref{fig:average}(a). Also, we have found no significant differences 
between either Tartarus spectra and those reduced independently (LZM01).

We then tested for the effects of various methods of averaging. When we
use arithmetic average instead of the (statistically) weighted one, we find the
peak of the line in the ratio plot increases, due to the contribution of
objects with a strong line measured with large errors. Since the dispersion in
our sample is dominated by statistical errors (over the intrinsic dispersion),
and the sample is {\it not\/} dominated by a few bright objects with small
statistical errors, the weighted average gives a much better representation of
the average properties. Still, even our unweighted line profile shows no
strong red wing (see LZM01 for details). On the other hand, neglecting the
weights due to different sizes of the bins has an effect of increasing the
scatter in the ratios.

We note that our results are not unexpected. There are relatively few Seyferts
with a confirmed strong red wing, the best cases being MCG --6-30-15 (Iwasawa
et al.\ 1996) and NGC 3516 (Nandra et al.\ 1999). In detailed studies of IC
4329A and NGC 5548, the lines were found to be weak and narrow, consistent with
the bulk of line formation relatively far away from the central black hole
(Done, Madejski \& \.Zycki 2000; Chiang et al.\ 2000). These findings are
clearly incompatible with the {\it average\/} line profile of Seyfert 1s being
virtually the same as that of MCG --6-30-15 (N97).

\section{PHYSICAL INTERPRETATION}
\label{s:slope}

Since the observed range of values of $\Gamma$ has to have a physical origin,
e.g.\ due to varying cooling rate of thermally-Comptonizing plasma
(e.g.\ ZLS99), we have ranked the spectra according to the spectral slope
obtained in fits excluding the line region and divided into 3 subsamples with
decreasing hardness (containing 22, 22 and 27 spectra, respectively). 
We then obtain 3 data-to-model ratios using the same method as above. An
immediately visible effect is that both the line strength and the strength of
the red wing increase with increasing softness of the continuum. To enable
spectral fitting, we multiply the ratios by the respective average power
law. Note that these spectra no longer use the \asca\/ response, and any intrinsically sharp feature is now broadened by the \asca\/ resolution. 

We have first found that the line profiles clearly contain 2 components, broad 
and narrow (similarly to the finding of Weaver et al.\ 1997 for MCG --5-23-16). 
The narrow line is well-modeled by a Gaussian with $\sfe^{\rm n}=0.1$ keV, 
corresponding to the time-averaged \asca\/ resolution. Thus, the {\it 
intrinsic\/} width of the narrow line is less than 0.1 keV. This indicates its 
origin at radii $\gg 10^3 GM/c^2$ (where $M$ is the black hole mass). The broad 
component originates, most likely, in an accretion disc. It is modeled as a 
line from a disc in the Schwarzschild metric (Fabian et al.\ 1989) with the 
inner radius, $r_{\rm in}$ ($\geq 6$; in units of $GM/c^2$) and the 
inclination, $i$, as free parameters. The outer radius is fixed at $10^3 
GM/c^2$, and we assume the line emissivity equal to the emissivity of a 
geometrically-thin disc. Since the fluorescent disc emission has to be 
accompanied by Compton reflection, we add it to the model (using Green's 
functions of Magdziarz \& Zdziarski 1995), also taking also into account its 
relativistic broadening treated in the same way as for the line (model {\tt 
refsch} in {\sc xspec} v.\ 11). The incident continuum is a power law e-folded 
with an energy of 300 keV (e.g.\ Gondek et al.\ 1996). We tie the reflection 
strength, $\Omega/2\upi$, to the equivalent width of the broad line, $\wfe^{\rm 
b}$ (\.Zycki, Done \& Smith 1998; Done et al.\ 2000), using results of George 
\& Fabian (1991). Specifically, for the obtained 3 best-fit values of $\Gamma$ 
and $i$ (see Table 1) we assume $\wfe^{\rm b}/(\Omega/2\upi)=150$, 135 and 120 
eV, respectively (thus, $\Omega/2\upi$ given in Table 1 is not an independent 
parameter). To take into account calibration uncertainty, we allow the 
local-frame line energy to be in the 6.35--6.45 keV range. Furthermore, since a 
structure around 7 keV is clearly seen in the average line profile (Fig.\ 
\ref{fig:average}), we include Fe K$\beta$ and Ni K$\alpha$ lines at 7.06 keV, 
7.46 keV with the photon flux equal to 0.12, 0.06, respectively, of the Fe 
$\alpha$ line, for both the broad and narrow component. We keep $i$ at the 
best-fitting value while determining the uncertainties of other parameters.

\begin{table*}
\centering
\caption{Results of fits to the 3 average \asca\/ spectra of Seyfert 1s
selected by their spectral hardness.  }
\begin{tabular}{cccccccc}
\hline
Spectrum & $\Gamma$ & $\Omega/2\pi$ & $i$ [deg] & $r_{\rm in}$ &
$\wfe^{\rm b}$ [eV] & $\wfe^{\rm n}$ [eV] & $\chi^2/\nu$ \\
\hline
a & $1.62^{+0.02}_{-0.03}$ & $0.48^{+0.12}_{-0.14}$ & $40^{+25}_{-7}$ &
$41_{-16}^{+22}$ & $72_{-18}^{+17}$ & $45_{-13}^{+12}$ &61.0/115 \\
b & $1.82^{+0.03}_{-0.03}$ & $0.88^{+0.23}_{-0.20}$ & $43^{+5}_{-4}$ &
$10_{-4}^{+14}$ & $115_{-27}^{+29}$ & $52_{-13}^{+11}$ &85.7/115 \\
c & $1.94^{+0.04}_{-0.03}$ & $1.00^{+0.21}_{-0.21}$ & $43^{+4}_{-1}$ &
$11_{-5}^{+5}$ & $124_{-26}^{+26}$ & $61_{-12}^{+11}$ &79.6/115 \\
\hline \end{tabular}
\end{table*}

\begin{figure}
\begin{center} \leavevmode \epsfxsize=7cm \epsfbox{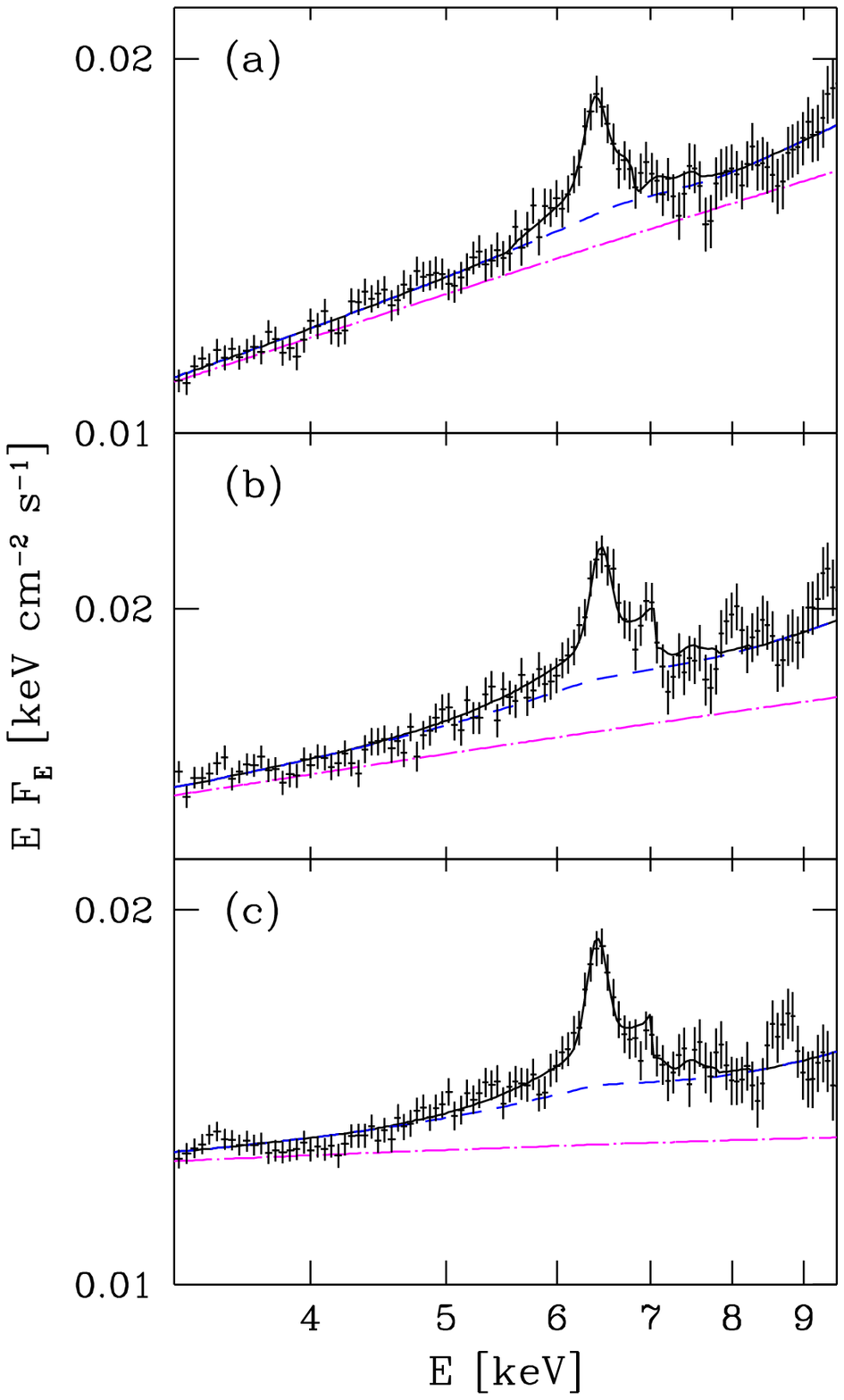} \end{center}
\caption{Average spectra of Seyferts (crosses) grouped according to the
decreasing spectral hardness (from top to bottom), fitted with the model (solid
curve) consisting of disc and narrow lines and Compton reflection, see Section
4 and Table 1. The dot-dashed and dashed curves show the incident power law and
the continuum including reflection. }
\label{fig:slope}
\end{figure}

The resulting spectral fits are shown in Fig.\ \ref{fig:slope}, and their
parameters are given in Table 1. We see that while $\wfe^{\rm b}$ and $\Gamma$
are clearly correlated, $\wfe^{\rm n}$ is practically constant (keeping
$\wfe^{\rm n}$ at the best-fit of the middle spectrum results in only $\Delta
\chi^2$ of $\sim 1$ for the other 2 spectra). Also, the inner radius of the
reflecting disc is much lower for the 2 softer spectra than for the hardest
one. This is consistent with a picture of cooling of the X-ray emitting plasma
by blackbody disc photons, with the same disc producing the broad Fe K line and
the Compton-reflected continuum (e.g.\ ZLS99). Then, the lower $r_{\rm in}$,
the softer the incident continuum (due to a decrease of the plasma temperature
caused by the increased cooling), and the more reflection and broad-line
emission. The fact that $r_{\rm in}\gg 6$ for our hardest spectrum is
consistent with either truncation of the cold disc and a hot inner disc
(Narayan \& Yi 1995), or, possibly, a complete ionization of the inner part of
the disc (Matt, Fabian \& Ross 1993) or a mildly relativistic outflow away from
the disc (Beloborodov 1999). The reflection region appears to be neutral as the
best fit of its ionization parameter is null in all three cases. Inclination is
consistent with being the same for all three spectra (as expected for $\Gamma$
independent of $i$), $i\simeq 43\degr$.

We have tested whether the obtained $\wfe$-$\Gamma$ correlation could be 
spurious. For that purpose, we have generated simulated spectra with a 
distribution of $\Gamma$ similar to that in our sample (see Fig.\ 
\ref{fig:individual}), the 3--10 keV flux equal to that in our averages, and a 
{\it constant\/} $\wfe$ of a relativistic line. Then, we obtained average 
spectra in 3 subsamples with the same ranges of $\Gamma$ as in our actual 
samples. Fits to the averages of the 3 simulated spectra reproduced the assumed 
$\wfe$ within 7\% with no systematic trend. This shows that the observed 
correlation is not spurious.

We have then checked that adding an unsmeared Compton-reflection component tied 
to the narrow line leads to a strong worsening of the fit, which implies that 
the remote medium is, on average, Thomson-thin (and thus does not produce 
noticeable Compton reflection, e.g.\ Wo\'zniak et al.\ 1998). The narrow line 
is present at a very high statistical significance, $>1-10^{-7}$ in all 3 
cases.

\begin{figure}
\begin{center} \leavevmode \epsfxsize=8cm \epsfbox{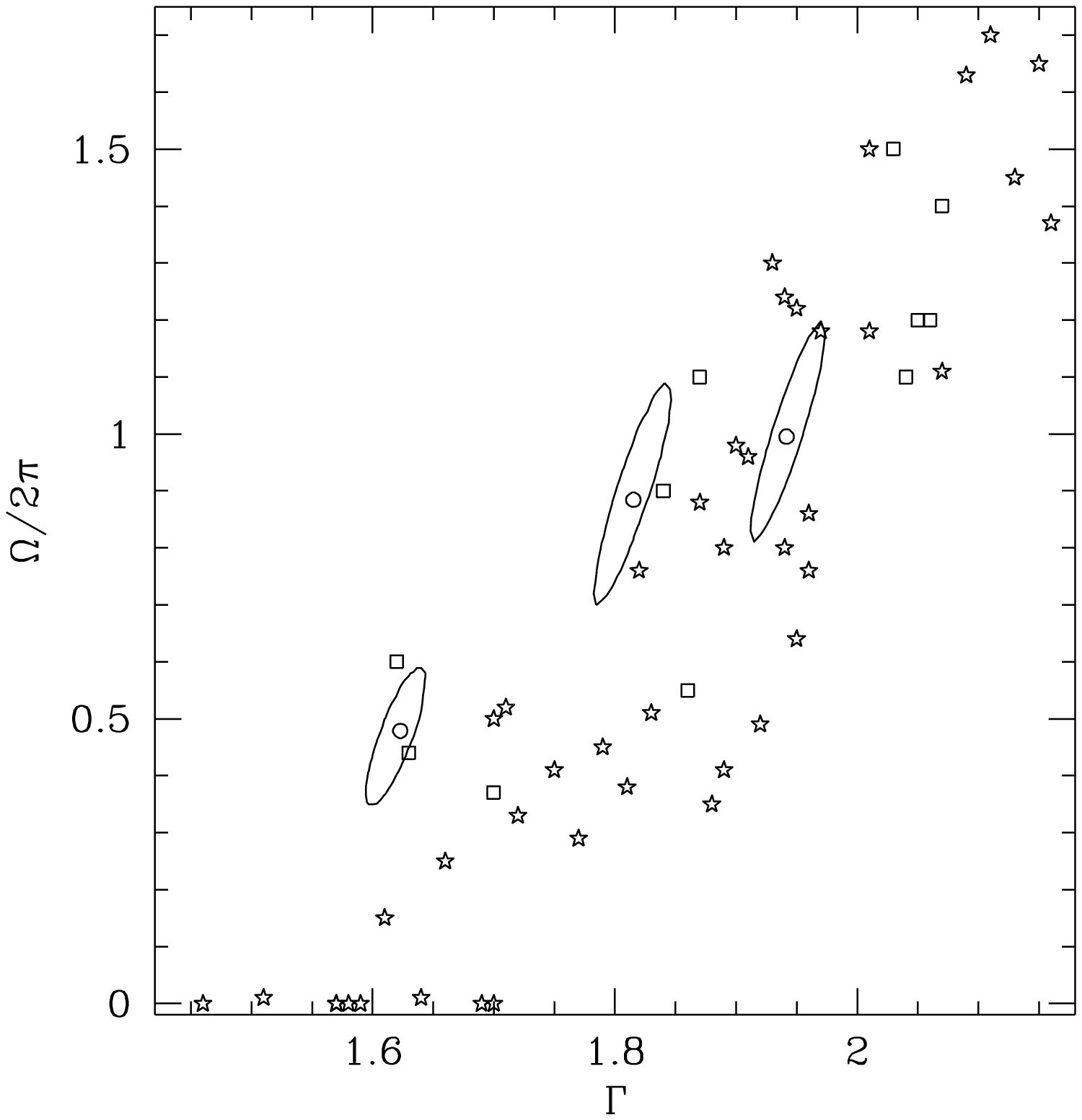} \end{center}
\caption{Relationship between the spectral slope and Compton reflection. The
present results are shown by the error contours (1$\sigma$, $\Delta\chi^2
=2.3$). Asterisks and squares are results from {\it Ginga\/} and {\it
BeppoSAX\/}, respectively. }\label{fig:correlation}
\end{figure}

We find that the $\Omega$-$\Gamma$ correlation seen in Table 1 is approximately 
consistent with that found in the data from \ginga\/ (ZLS99) and {\it 
BeppoSAX\/} (Matt 2001), see Fig.\ \ref{fig:correlation}. The shown best-fit 
\ginga\/ points have been obtained here by refitting the data of ZLS99 with our 
present physical model (at $i=43\degr$ and $r_{\rm in} (\Gamma)$ as in Table 
1). Some offsets in the values of $\Omega/2\upi$ and $\Gamma$ between the 
\ginga, \sax\/ and \asca\/ results are apparently due to calibration 
differences between the instruments. Still, we see that our present 
$\Omega$-$\Gamma$ correlation (based on the Fe K line) is in good agreement 
with those driven by the strength of the Compton reflection.

We note that Nandra et al.\ (2000) found that some systematic effects 
(background undersubtraction with a specific power-law dependence on energy and 
time lag of reflection) may lead to an apparent correlation between $\Omega$ 
and $\Gamma$ in a {\it single\/} AGN with constant $\Omega$. However, time-lag 
effects average out in a sample of AGNs. Then, it is highly unlikely that 
background-subtraction effects are responsible for the correlations in the AGN 
samples observed by all \ginga, \sax\/ and \xte\/ (and now confirmed by \asca). 
They also found that statistical effects alone cannot explain the correlation 
observed even in a single object (NGC 7469), in full agreement with the 
analysis of ZLS99. On the other hand, Vaughan \& Edelson (2001) have shown that 
an apparent $\Omega$-$\Gamma$ correlation will be seen in multiple observations 
of a single AGN with these quantities being constant. However, this is just 
equivalent to showing an error contour for measurements of the 2 quantities, 
which is elongated and skewed (ZLS99). This effect was taken into account by 
ZLS99, who computed the  probability of the correlation {\it beyond\/} that due 
to that error contour appearing by chance to be $\sim 10^{-10}$ in the \ginga\/ 
sample. Such a low probability is due most of the error contours in ZLS99 being 
much smaller than the extent of the correlation. This is in contrast to the 
case considered by Vaughan \& Edelson (2001), in which the extent of their 
apparent correlation is equal to the length of the measurement error contour. 

Note that a very strong  $\Omega$-$\Gamma$ correlation is also found in 
accreting stellar-mass black holes. In that case $\Omega$ also correlates with 
$\sigma_{{\rm K}\alpha}$ and with the characteristic frequencies in the power 
spectrum (Gilfanov et al.\ 1999, 2001). 

As discussed above, the $\Omega$-$\Gamma$ correlation appears, most likely, due 
to feedback between hot and cold media. Its presence is of major importance as 
a diagnostic of geometry and physical conditions in accretion flows (e.g.\ 
Beloborodov 2001). One of its consequences is ruling out a geometry with static 
active regions above a disc, as it predicts an {\it anticorrelation\/} between 
$\Omega$ and $\Gamma$ (Malzac, Beloborodov \& Poutanen 2001). 

\section{CONCLUSIONS}

We have studied Fe K line properties of a large sample of Seyfert 1s observed
by \asca. Gaussian fits yield $\langle \sfe\rangle = 0.22\pm 0.03$ keV and
$\langle \wfe\rangle= 0.13\pm 0.01$ keV, i.e., relatively weak and narrow
lines. The average line profile of our sample consists of a narrow core and red
and blue wings. The average red wing is much weaker than that of the Seyferts
with broadest K$\alpha$ lines, e.g.\ MCG --6-30-15.

Since the power-law slope of Seyferts in our sample covers a large range, we
have divided our sample into 3 groups, which average spectra we fitted
separately. All 3 spectra contain a narrow core with $\wfe^{\rm n}\simeq 50$
eV, which we interpret as originating from a Thomson-thin remote medium. In
addition, the data show a broad feature, well-fitted by a line from a
relativistic disc. The equivalent width of the broad line increases from
$\wfe^{\rm b} \sim 70$ eV to $\sim 120$ eV with increasing $\Gamma$. The fitted
inner disc radius decreases from $r_{\rm in}\simeq 40$ for $\Gamma\simeq 1.6$
to $r_{\rm in}\simeq 10$ at $\Gamma\sim 1.8$--2. The inclination is consistent
with being $\sim 45\degr$ on average. The correlation of $\wfe^{\rm b}$ with
$\Gamma$ corresponds to a correlation between the strength of Compton
reflection and $\Gamma$, confirming previous results of ZLS99 and Matt (2001).

Our results resolve the long-standing discrepancy (see \S 1) between the 
properties of the Fe K lines and those of Compton reflection in Seyfert 
galaxies. 

\section*{ACKNOWLEDGMENTS}

We thank M. Gierli\'nski for help with {\sc xspec}, C. Done, K. Leighly and 
G. Madejski for valuable discussions, and the anonymous referee for careful 
checking the results of this paper. This research has been supported in part 
by a grant from the Foundation for Polish Science and KBN grants 2P03D00614 and 
2P03C00619p0(1,2), and it has made use of data obtained through the High Energy 
Astrophysics Science Archive Research Center Online Service, provided by 
NASA/GSFC.


\label{lastpage}

\end{document}